\documentclass[review,number,sort&compress]{elsarticle}

\usepackage{lineno,hyperref}
\modulolinenumbers[5]
\usepackage{graphicx}
\usepackage{color}
\journal{Nuclear Instruments and Methods in Physics Research A}

\newcommand{\rootfigscale}{0.4}
\begin{document}

\begin{frontmatter}

\title{A gas cell for stopping, storing and polarizing radioactive particles}

\author{A.~Sytema\corref{correspondingauthor}}\ead{A.Sijtema@rug.nl}
\author{J.E.~van~den~Berg}
\author{O.~B\"oll}
\author{D.~Chernowitz}
\author{E.A.~Dijck}
\author{J.O.~Grasdijk}
\author{S.~Hoekstra}
\author{K.~Jungmann}
\author{S.C.~Mathavan}
\author{C. Meinema}
\author{A.~Mohanty}
\author{S.E.~M\"uller\fnref{Mfootnote}}
\fntext[Mfootnote]{Present address: Helmholtz-Zentrum Dresden-Rossendorf, Germany.}
\author{M.~Nu\~nez Portela}
\author{C.J.G.~Onderwater}
\author{C.~Pijpker}
\author{L.~Willmann}
\author{H.W.~Wilschut}

\address{Van Swinderen Institute for Paricle Physics and Gravity, University of Groningen, Nijenborgh 4, 9747AG Groningen}
\cortext[correspondingauthor]{Corresponding author}

\begin{abstract}
A  radioactive beam of $^{20}$Na is stopped in a gas cell filled with Ne gas. The stopped particles are polarized by optical pumping.
The degree of polarization that can be achieved is studied. A maximum polarization of 50\,\% was found.
The dynamic processes in the cell are described with  a phenomenological model.
\end{abstract}


\end{frontmatter}


\section{Introduction}
Particles of a radioactive secondary  beam can be  stopped in  high-pressure buffer gas, where they can be trapped sufficiently long to study their decay, if their lifetime is shorter or of order of the typical diffusion time. If the beam particles neutralize to atoms, the nuclei of certain elements can be spin-polarized by optical pumping. This method for studying polarized particles was developed by Otten and coworkers~\cite{Schweickert1975} and more recently employed by Backe et al.~\cite{Backe1992}, Young et al.~\cite{Young1995}, and Voytas et al.~\cite{Voytas1996}. We have exploited it to study Lorentz violation in weak interactions, i.e. the question whether the integrated $\beta$-decay rate depends on the nuclear-spin direction with respect to an absolute reference system~\cite{Muller2013, Sytema2015}. The analysis of these experiments required an effective description of the time dependence of the polarization, which will be discussed in this article. 

The measurements were made with  a beam of $^{20}$Na stopping in Ne buffer gas.  
By reversing the polarization direction and by switching the particle beam on and off,  the characteristic time dependence of the polarization can be measured and the dynamics in the gas cell can be inferred to some extent. The relatively short half-life of $^{20}$Na of 0.45\,s is essential for these measurements. 

The polarization technique used here requires neutralization of the incoming beam.  Therefore, the experimental situation is analogous to gas catchers operated to extract secondary ion beams~\cite{Wada2013,Marsh2014}. We discuss the  operation of our gas cell in this context.
A difference is that in this experiment stable Na can be dispensed in the gas cell.

The gas cells discussed here operate with noble gas;  the  ionization potential of the noble gas element is higher than that of the incoming particle such that it can remain singly charged. However, to which extent the particles in a fast ion beam are  neutral, once they thermalize in a gas cell,  remains an open question.  If they remain ions, they can  be extracted by flowing the buffer gas out. Ion catchers operate on this basis, a recent review is in Ref.~\cite{Wada2013}. If the particles neutralize, they can be re-ionized by two-step laser resonance ionization, which is element selective. A recent review of such Laser Ion Sources (LIS) is in Ref.~\cite{Marsh2014}.  

 A detailed study of neutralization in LIS is in Refs.~\cite{Kudryavtsev2001,Facina2004}. Also the role of chemical binding of the stopped ion with trace molecules in the buffer gas has been discussed there. Here we show that by dispensing natural Na we appear to bind those molecules that otherwise would bind $^{20}$Na. We find that the polarization maximizes to about 50 \% and the polarization lifetime of $^{20}$Na is a few seconds.

The outline of this paper is as follows. We describe the experimental setup and the basic experimental observations. To facilitate the discussion of our results the time scales of various relevant processes in the gas cell are quantified.  Next, our experimental results are presented and a phenomenological description is given. We conclude with a summary of our findings.

\section{Experimental setup}
\begin{figure}\centering
\includegraphics[width=0.5\textwidth]{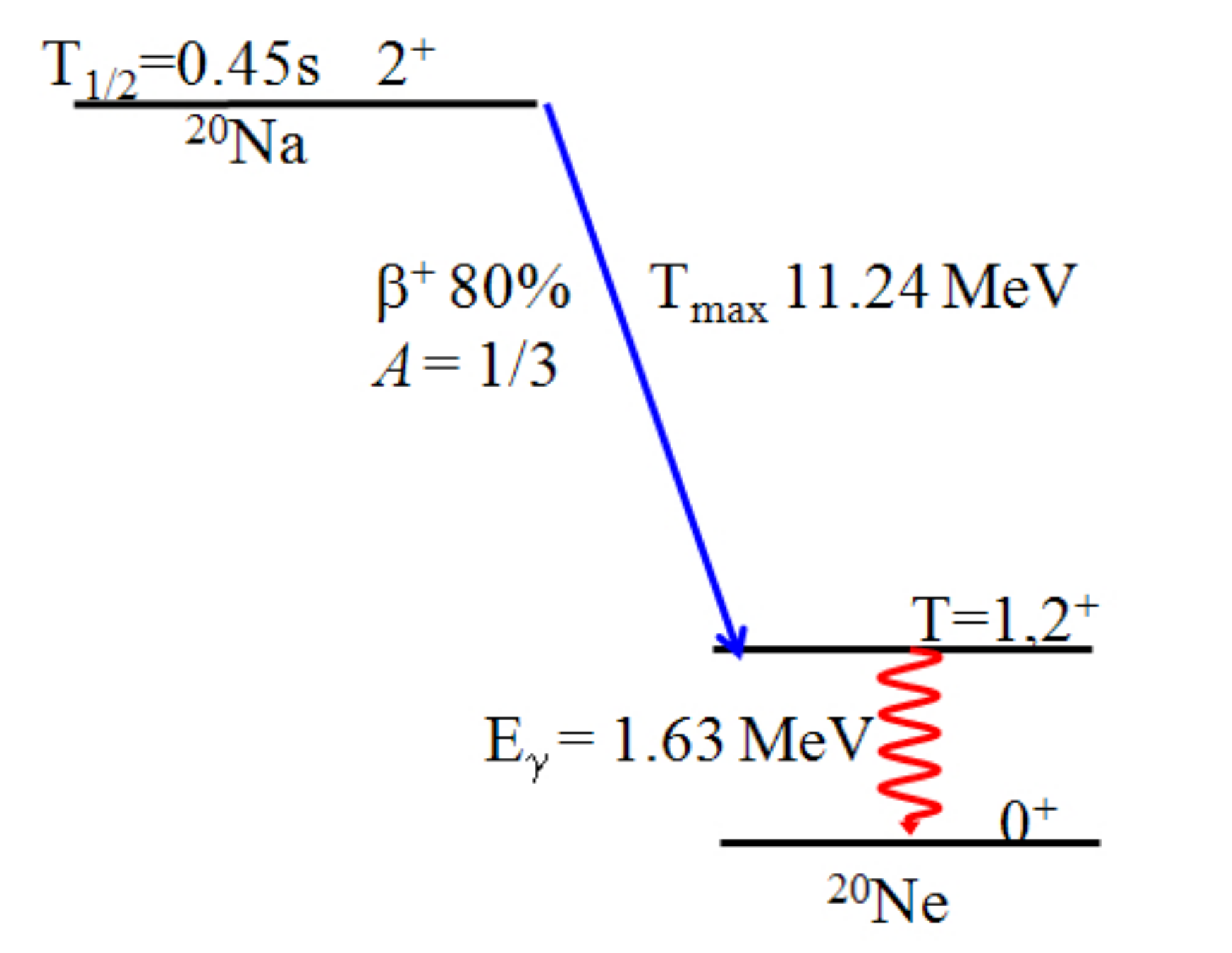}
\caption{Relevant decay properties of $^{20}$Na. }\label{levelscheme}
\end{figure}

\begin{figure}
\centering
\includegraphics[width=0.75\linewidth]{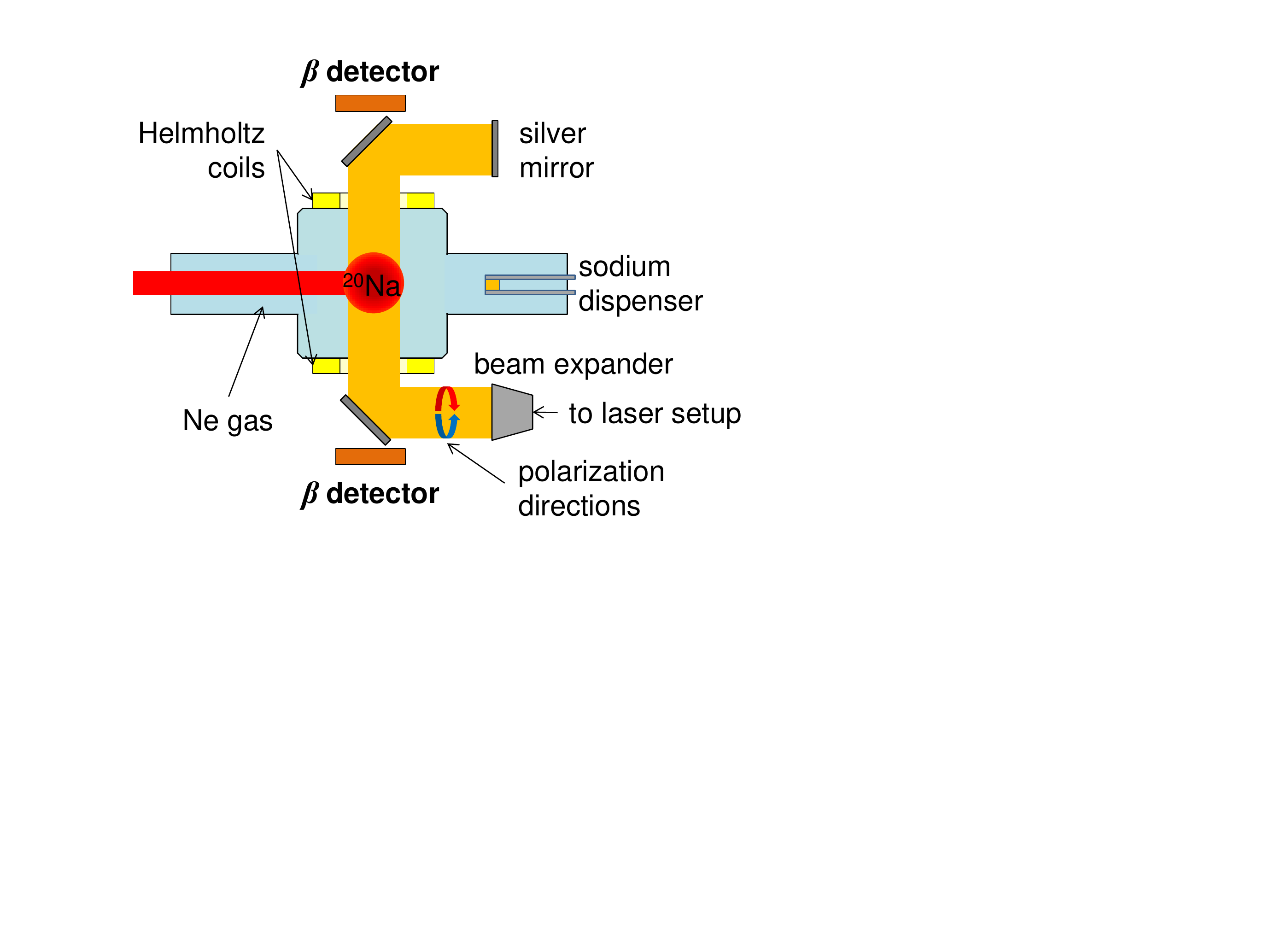}
\caption{Schematic drawing of the buffer-gas cell, the $^{20}$Na beam, the light beams, and detectors.}
\label{setupfig}
\end{figure}

We used the secondary $^{20}$Na beam from the TRI$\mu$P facility in Groningen \cite{Traykov2007}. The relevant nuclear properties of $^{20}$Na are displayed in Fig.~\ref{levelscheme}.
The energetic beam of approximately 20 MeV/nucleon entered through a metallic foil into a gas cell filled with neon at 7 atm absolute pressure.
The schematic setup is shown in Fig.~[\ref{setupfig}].
The fiducial part of the setup is made from a stainless steel cube with $70\,\mathrm{mm}$ sides and $38\,\mathrm{mm}$ bore diameter in the three principal directions.
The cell was filled with neon through a liquid-nitrogen cooled trap and  a gas purifier.
Natural sodium can be evaporated into the cell by heating a commercially available sodium dispenser.
Aluminum foils were placed in front of the cell such that the beam stops in the middle of the fiducial volume. The position fine-tuning was made by appropriately rotating one of the foils for maximizing the count rate in the detectors. 

The laser beam was tuned to the $^{20} \mathrm{Na}$ $D_1$ transition adjusted for the buffer gas pressure ($\lambda = 589.782\,\mathrm{nm}$).
Pressure broadening of about $50\,$GHz mixes the hyperfine states.
 The laser beam is split into two beams with opposite light helicity and recombined again onto the same optical path. Actuators allow to select the light of either beam or to block both beams. A beam expander magnifies the profile of the laser beam to an approximately Gaussian shape with $1.2\,$cm full-width-half-maximum.
 This laser beam is guided by silver mirrors and passes through fused silica windows (diameter $29\,\mathrm{mm}$) of the gas cell.
The average laser-light intensity delivered to the gas-cell fiducial area is $s_A=2\times 10^{-2}\,\mathrm{W}\,\mathrm{cm}^{-2}$.
The windows are surrounded by coils in Helmholtz configuration for a magnetic field of about $15\,$Gauss aligned with the laser beam.

The $\beta$ particles from $^{20}$Na decay have an endpoint energy of 11.7 MeV. $\beta$ particles with energy exceeding about $2\,$MeV can pass through the window and the mirror. The average velocity of the detected $\beta$ particles is 99\% of the light speed.
They are measured with two thin scintillation detectors, that are insensitive to $\gamma$ rays. The $\beta$ detectors cover an angular range $\Delta \theta=11^\circ$ relative to the polarization direction, i.e.~$\langle \cos\theta\rangle=0.99$.

\begin{figure}
\centering
\includegraphics[scale = \rootfigscale]{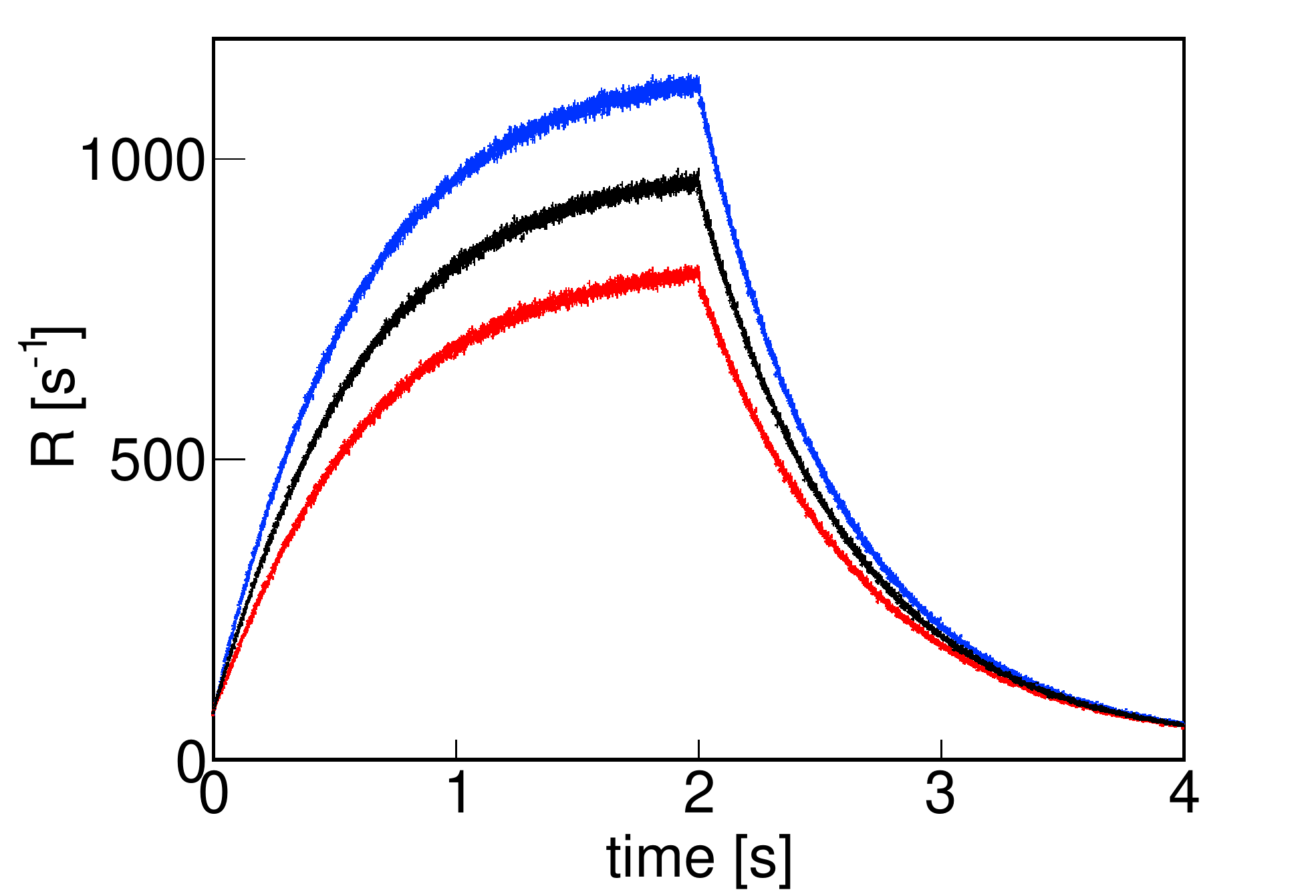}
\caption{The measured $\beta$ rates in one of the detectors. The top (blue) and bottom (red) curves are data points obtained with opposite helicity of the laser light. The middle curve displays the decay rate when the laser light was absent.}
\label{beta}
\end{figure}

Figure~\ref*{beta} displays the $\beta$-particle rates measured in periods with the primary $^{20}$Ne beam on ($0-2\,$s) and  off ($2-4.1\,$s).
In three consecutive periods, the samples were first polarized in one direction ($0-4\,$s), then in the opposite direction ($4.1-8.1\,$s),
and in the third period ($8.2-12.2\,$s) the laser light was off. The additional $0.1\,$s was required for operation of the actuators, here the polarization is undefined.
The data of many sequences have been averaged here to obtain good statistical accuracy for each $4\,$s period.

The $\beta$-asymmetry parameter $A_\mathit{Wu}=1/3$ for a $2^+ \rightarrow 2^+$ Gamow-Teller transition.
Therefore,  to good approximation, the count rate in the $\beta$-particle detectors is 
$R_{L(R)}^\pm \propto 1\mp(\pm) A_\mathit{Wu} P $, 
where the signs depend on the direction of the polarization ($\pm$) and the place (\textit{L/R}) of the detector.
The polarization is  $P=A_\beta/A_\mathit{Wu}$.
The count rate asymmetry $A_\beta$ is given by 
\begin{equation}\label{eq:Abeta}
A_\beta = \frac{\sqrt{R_L^+R_R^-}-\sqrt{R_L^-R_R^+}}{\sqrt{R_L^+R_R^-}+\sqrt{R_L^-R_R^+}} \;. 
\end{equation}
\begin{figure}
\centering
\includegraphics[scale = \rootfigscale]{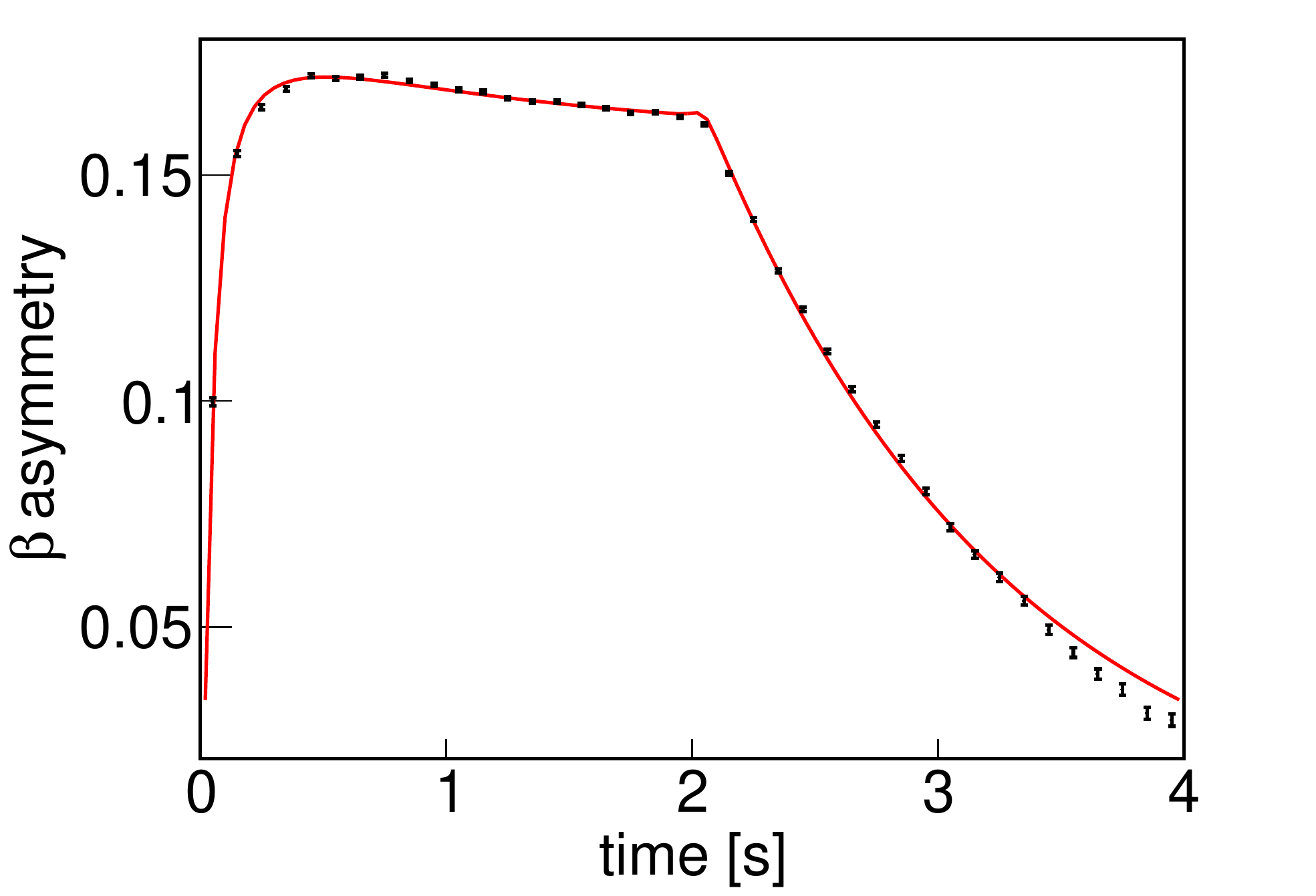}
\caption{The $\beta$ asymmetry $A_\beta$ averaged over time periods of 4 s.  The curve through the data points is discussed in Section~\ref{experimental}.}
\label{polall}
\end{figure}
The observed value corresponding to the data in Fig.~\ref{beta} is shown in Fig.~\ref{polall}.
The maximum polarization of about 50\% is achieved when the beam is on, while it drops when the beam is off.
This observation is central to the discussion in Section \ref{experimental}.

\section{Characteristic time parameters}
For our discussion  it is useful to consider the time scale of various processes that occur if fast ions are stopped in the gas cell, neutralize and are polarized.

\subsection{Neutralization}\label{neutralization}
There could be various scenarios for neutralizing the incoming beam. The stopping beam will ionize the Ne gas. Ionization is maximal in the region of the Bragg peak. Beyond the Bragg peak, electronic stopping will not be dominant anymore, instead the particle suffers atomic collisions by which it reaches thermal energies. In the following we list possible processes leading to neutralization, and give an estimate of their characteristic time and importance, when possible. Current knowledge is insufficient to establish their relative relevance.

\subsubsection{Neutralization in stopping} The stopping particle will slow down eventually reaching the electronic stopping regime where it neutralizes (cross section $\sigma_{10}$) and re-ionizes (cross section $\sigma_{01}$), such that its average charge state is $\langle q \rangle=\sigma_{10}/(\sigma_{10}+\sigma_{01})$ \cite{Allison1958}.
In general, the cross-section values are not known in the energy  regime where electronic stopping ends and collisional stopping, not involving charge transfers, starts.
 \subsubsection{Neutralization in a uniform plasma}
 When the incoming ion intensity is sufficiently high a plasma is formed in the buffer gas.
 Assuming the density of ions and atoms to be $n=n_\mathrm{ion}=n_\mathrm{electrons}$, neutralization is the result of a three-body process $\mathrm{X}^+ +\mathrm{e}^- +\mathrm{Ne}\rightarrow \mathrm{X}+ \mathrm{Ne}$, where X can be either a Ne or Na ion.
Following Section 3.1 in Ref.~\cite{Kudryavtsev2001} one has
\begin{eqnarray}\label{eq:uniform}
\frac{dn}{dt}&=&Q-\alpha n^2 ,\\
n(t)&=&\sqrt{\frac{Q}{\alpha_3}}\tanh(t\sqrt{Q\alpha_3}).\nonumber
\end{eqnarray}
Here $Q=f_i I_\mathrm{Na}$ is the average density of ion-electron pairs created per second due to a  beam of $I_\mathrm{Na}$ particles/s. In our experiment
$f_i\approx 10^{6}\,\mathrm{cm}^{-3}$.
This value is based on  the longitudinal spread in  beam energy ($\pm 3\%$, corresponding to $1.5\,$cm) 
and lateral extension (determined by an up-stream collimator with an area of  $3\,$cm$^2$). 
We have used the SRIM code \cite{SRIM2008} to obtain the energy loss in terms  of electron ion pairs, using $36\,$eV per pair \cite{Sipila1976}. The main contribution to the electron-ion density  arises from the Bragg peak.  The constant $\alpha_3$ is the three-body recombination coefficient, which has been measured for Ne to be of order $10^{-5}\,\mathrm{cm}^{3}\mathrm{s}^{-1}$~\cite{Dolgovsa1970}.
Therefore, the equilibrium density of $n=\sqrt{f_i I_\mathrm{Na}/\alpha_3}\approx 10^8\,\mathrm{cm}^{-3}$ is reached 
in a characteristic time $\tau_n=1/\sqrt{f_i I_\mathrm{Na}\alpha_3}\approx 0.3\,$ms. When this scheme applies, nearly all Na ions can be neutralized  assuming  for Na the same recombination coefficient as for Ne.
When the beam is turned off the  ion and electron density decreases as $1/(1+t/\tau_n)$, dropping several orders of magnitude within a fraction of a second.

\subsubsection{Neutralization by Bragg peak electrons}
The assumption of a uniform plasma  only holds for a high intensity of incoming particles.
Therefore, we consider also individual events. The stopping particle moves beyond the region where the Bragg maximum is located (here a distance $\delta \approx 2\,$mm).  The local plasma can expand by diffusion while at the same time electrons and Ne ions recombine. To study its relevance for the present situation, we made a schematic calculation by modifying Eq.~(\ref{eq:uniform}) as follows
\begin{eqnarray}\label{eq:notuniform}
\frac{\partial n(t,r)}{\partial t}&=&D \triangle n(t,r) - \alpha n^2(t,r) ,\\
\frac{ d q_{\mathrm{Na}}(t)}{dt}&=&1 - \alpha n(t,\delta) q_{\mathrm{Na}}(t),\nonumber
\end{eqnarray}
where the boundary condition $n(0,r)$ is an appropriate Gaussian at the position of the Bragg maximum. We used a spherical Gaussian with width $\sigma_r=0.13\,$cm containing $10^6$ ion-electron pairs describing the ionization distribution at the Bragg maximum.
 $D=0.014\,\mathrm{cm^2/s}$ is the diffusion constant of  Ne ions in $7\,$atm Ne \cite{Jovanovic2002}. The electron density distribution is assumed to be the same as that of the ion, keeping the local plasma neutral. With these approximations, the average charge state  $\langle q_{\mathrm{Na}} \rangle$ develops with a characteristic time of about $50\,$ms to a minimum of 0.1, i.e. $\approx 90\,\% $ neutralization. The result depends strongly on the initial conditions, but shows that it is a viable option for neutralization, which has hitherto not been considered in the literature. 

\subsubsection{Re-ionization}
We also need to consider the re-ionization of Na atoms by Ne ions.
In  Langevin's approximation this rate can be calculated \cite{Gioumousis1958}; 
it is given by $\langle\sigma v\rangle=2\pi\sqrt{(e^2\alpha_p/\mu)}$, 
where $\alpha_p=163\,\mathrm{a.u.}$ is the polarizability of Na \cite{Thakkar2005} and $\mu$ the reduced mass.
This evaluates to $\langle\sigma v\rangle \approx 10^{-9}\,\mathrm{cm^3s^{-1}}$.
Because this means that $\langle\sigma v\rangle << \alpha_3$, re-ionization by Ne ions does not play a role.

\subsubsection{Resonant charge transfer}\label{sct:resonant}
An alternative to neutralization in absence of a plasma, is the use of resonant charge transfer between stable and radioactive Na. The cross section for charge exchange is about $5\times10^{-14}\,\mathrm{cm^2}$ \cite{Oluwole1977,Nienstadt1978,Agagu1980}, which gives  $\langle\sigma v\rangle =3\times 10^{-9}\,\mathrm{cm^3s^{-1}}$. Neutralization can then be achieved at a $1\,\mathrm{ms}$ rate using Na at a vapor pressure of $10^{-8}\,\mathrm{atm}$ corresponding to a temperature of $150^\circ\,\mathrm{C}$. The present system required operation at room temperature where the vapor pressure is more than four orders of magnitude lower. Therefore, resonant charge exchange is not a relevant mechanism in the current setup. In other experiments \cite{Young1995,Voytas1996}  the relevant natural alkali elements  was also added and the setup was kept at elevated temperature. The polarization in these experiments was nearly 100\,\% indicating that complete neutralization had been achieved.

\subsection{Optical pumping}\label{pumping}
Assuming neutralized Na, there are two time constants characterizing the polarization.
These can be associated with optical pumping leading to a stretched configuration in the magnetic substate distribution 
while on the other hand collisions in the gas destroy this distribution.
The two times $\tau_1$ and $\tau_2$, respectively can be calculated using a model based on the formalism of Happer \cite{Happer2010}.
The model assumes a depolarizing cross-section of the ground state of
$1.8\times 10^{-23}\,\mathrm{cm}^2$~\cite{Walker}
and the depolarizing cross section for the $J = 1/2$ excited state of
$2.9\times 10^{-15}\,\mathrm{cm}^2$~\cite{Gay1976}.
We find $\tau_1 = 9.3\times 10^{-4}\,$s and $\tau_2 = 5.9\times 10^{-2}\,$s.
$\tau_1$ is inversely proportional to the laser intensity,
while $\tau_2$ is inversely proportional to the ground state depolarization cross section.
The maximum polarization possible is $\tau_1/(\tau_1+\tau_2)=0.98$, and should be achieved with a characteristic time 
$\tau_1 \tau_2/(\tau_2+\tau_1)=9.2\times 10^{-4}\,$s.

\subsection{Diffusion}\label{sect:diffusion}
Na atoms may diffuse out of the laser light or can be absorbed on the container walls before decaying.
The characteristic displacement of the stopped atoms is $l=\sqrt{6 D\tau}$.
The diffusion of Na in Ne was measured \cite{Voytas1996} as $D=0.31\,\mathrm{cm^2s^{-1}}$ at STP. Then, with $7\,$atm Ne, $l=0.4\,\mathrm{cm}$ to be compared with a fiducial volume of length scale $\sim 2\,$cm.
One can show that in the present  setup, an initial spherical source with a full-width-half-maximum of $1.5\,$cm (approximately the stopped ion distribution) diffuses so little that a polarization lifetime of $> 10\,\mathrm{s}$ is possible.  Therefore, diffusion is not expected to play a role of importance.

\subsection{Contaminants}\label{sct:contaminants}
Contaminants will bind atomic Na, thereby also reducing the polarization lifetime. A typical reaction rate is $10^{-30}\,\mathrm{cm^6s^{-1}}$ for e.g.~$\mathrm{Na} + \mathrm{O_2}+ \mathrm{Ne}\rightarrow\mathrm{NaO_2}+\mathrm{Ne}$ ~\cite{Fontijn2013}. This would require a concentration of $\mathrm{O_2} < 0.3\times 10^{-12} $ in order to achieve a lifetime of free Na atoms beyond one second. Dispensing an  appropriate alkali element thus reduces active contaminants. If a high vapor pressure can be maintained it serves a dual purpose by also aiding in neutralization (see section~\ref{sct:resonant}).

\section{Experimental results}\label{experimental}
\begin{figure}
\centering
\includegraphics[scale = \rootfigscale]{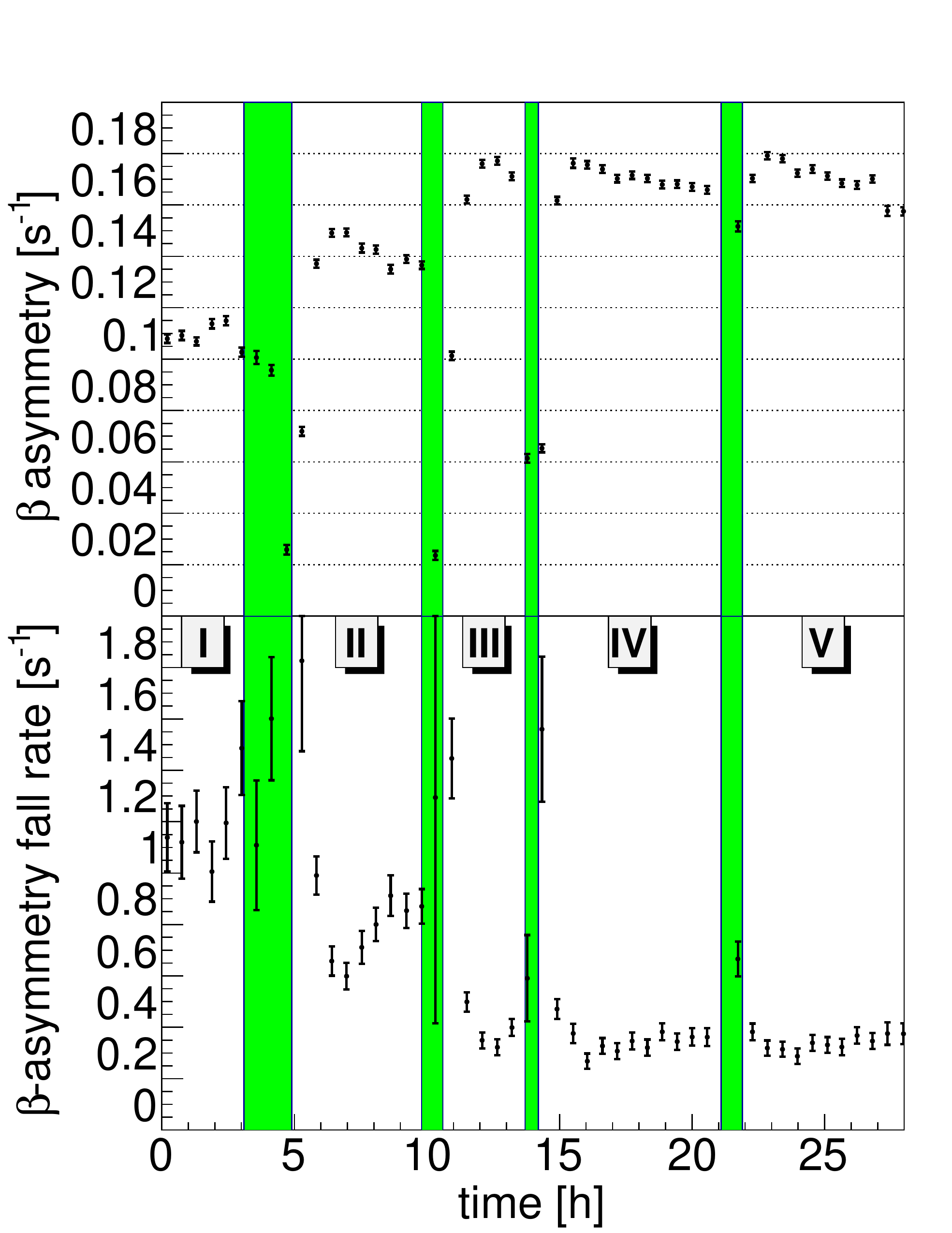}
\caption{The effect of adding natural Na to the gascell. Shown is the evolution of the $\beta$ asymmetry (cf.~ Fig.~\ref{beta}) during some 24 hours: the average asymmetry in the region $0.8<t<1.8$~s  and the  rate by which the asymmetry decreases when the beam is stopped $t>2$~s, assuming an exponential drop.
In the colored regions  the Na dispenser was turned on.}\label{fig:getter}
\end{figure}
The main time-dependent features of $A_\beta$ (Fig.~\ref{polall}) are the maximum asymmetry (polarization) after the radioactive beam is entering the cell and the subsequent exponential polarization loss when the beam is off.
At the start of the experiment very small polarization was observed. After adding natural Na dispensed from a getter, polarizaton was found. A stepwise increase of the polarization was observed after heating the Na dispenser. In Fig.~\ref{fig:getter} the maximum polarization and the polarization-loss rate are shown as function of time. The colored regions indicate the periods where the dispenser was turned on. During that time convection in the cell is large with the consequence that the polarization decreases and  the loss rate increases. However, after the  cell  returns to room temperature the cell performance improved. This procedure is repeated until no further improvement can be observed. 

Because resonant charge exchange does not play a role at room temperature (see section~\ref{sct:resonant}), we conclude that the natural Na binds the contaminants (see section~\ref{sct:contaminants}), that otherwise would bind radioactive Na. 
While it appears that no further dispensing of the stable isotope $^{23}$Na is necessary, we find that over several days the performance decreases: the maximum polarization decreases by 12\,\% over a period of 100 hours. The polarization-loss rate in the beam-off periods increases, first it doubles in  30 hours and then goes up with 12\,\% in the next 70 hours. 

We have modeled the time dependence of the polarization to include the various processes in the cell.
Because diffusion is not important, this can be achieved by solving a set of coupled differential equations that describe (1) the neutralization of incoming ions with a characteristic time $\tau_n$,
(2) the gain and loss term in the pumping to the polarized state given by $\tau_1$ and $\tau_2$ respectively,
and (3) two loss terms given by $\tau_\mathrm{loss1}$ when the beam is on, and $\tau_\mathrm{loss2}$ when the beam is off. 
The term "loss" means here that Na particles can not be polarized any longer, but they remain in the fiducial volume of the $\beta$-particle detectors, thus lowering the observed polarization. We find that it is not well possible to distinguish $\tau_n$ and $\tau_1$ in this description. For that reason we put $\tau_n$ at zero and vary $\tau_1$. We also keep $\tau_2$ at the calculated value of $60\,$ms.  A good description of the asymmetry is then obtained with three parameters, i.e. $\tau_1$, $\tau_\mathrm{loss1}$, and $\tau_\mathrm{loss2}$ as shown in Fig.~\ref{beta}. The parameters in this figure are
$\tau_1=41.3\pm0.5\;\mathrm{ms}$, $\tau_\mathrm{loss1}=3.9\pm0.2\;\mathrm{s}$, and $\tau_\mathrm{loss2}=1.22\pm0.01\;\mathrm{s}$. The description is sufficiently accurate for the purpose of the analysis in \cite{Sytema2015}.
\begin{figure}
\centering
\includegraphics[width=0.75\linewidth]{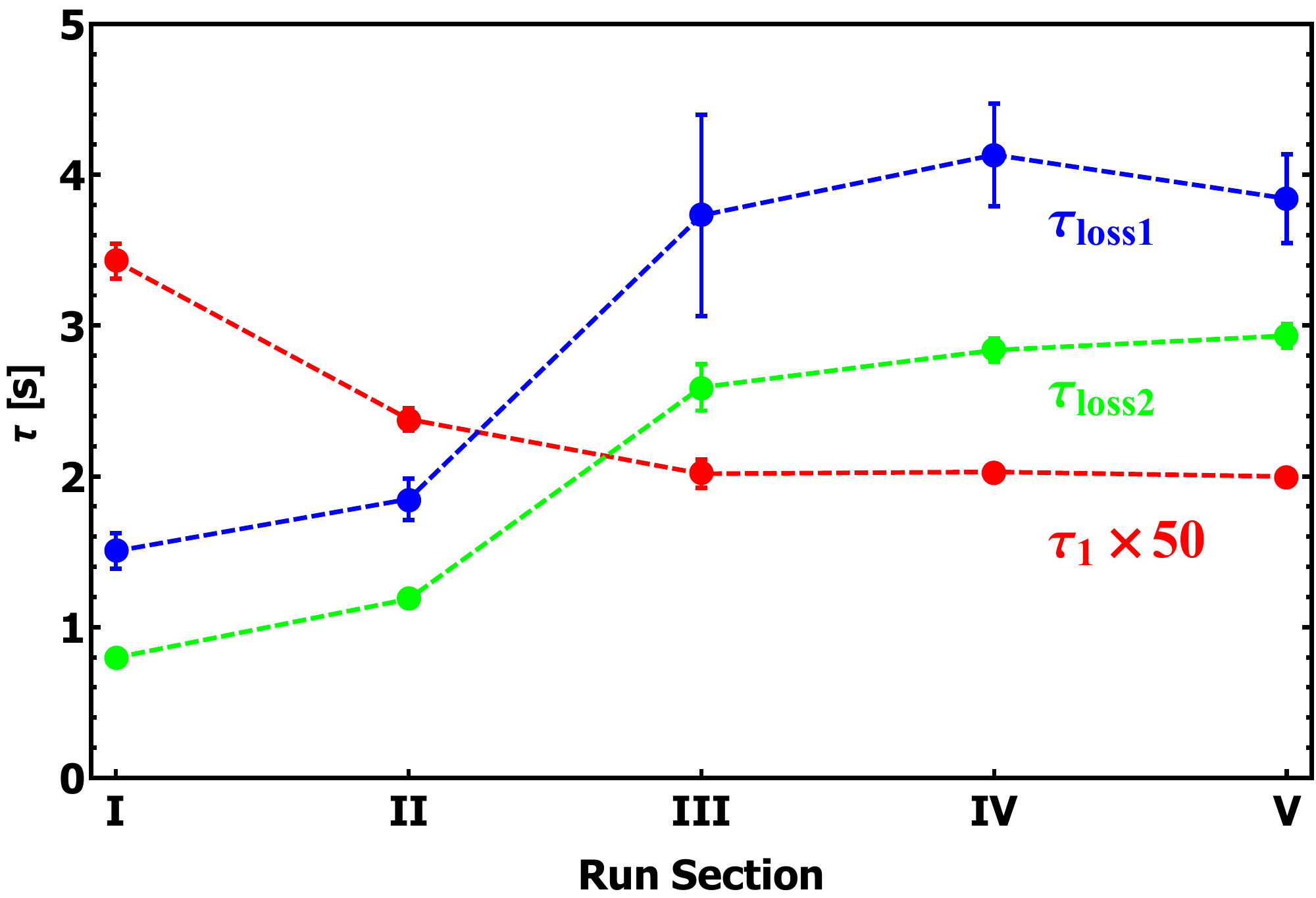}
\caption{Evolution of the three time parameters for the five episodes after dispensing natural sodium as shown in Fig.\ref{fig:getter}. The connecting lines are to guide the eye. }\label{fig:fitresults}
\end{figure}
Note that with this choice of parametrization $\tau_1$ determines both the rise time and magnitude of the polarization (cf.~section~\ref{pumping}), $P=59.2\pm0.3\;\%$.
However, the value of $\tau_1$ is determined mainly by the degree of polarization.  The time resolution due to the actuators  switching between polarized beams is of the same order as $\tau_1$. Therefore, although it appears that the rise time in the polarization is well described, this is probably accidental. Nonetheless, one can conclude that neutralization and polarization take place at a time scale smaller than or equal to the fitted value $\tau_1$. In Fig.\ref{fig:fitresults} we find the largest lifetimes $\tau_\mathrm{loss1}\approx 4\,\mathrm{s}$ and $\tau_\mathrm{loss2}\approx 3\,\mathrm{s}$. The longer lifetime in the beam-on period is ascribed  to the presence of ions and electrons, binding the residual contaminants.
The long lifetimes $\tau_\mathrm{loss1,2}$ show that the concentration of contaminants that bind Na are extremely low (cf.~Section~\ref{sct:contaminants}). 

The reason why the polarization is not maximal remains yet unclear. Either neutralization is incomplete or the range distribution is much wider than calculated, exceeding the diameter of the laser beam. In all gas cells, even in the absence of a drift field \cite{Facina2004} charged stopped ions were always observed. We also note that in principle neutralization can be studied using switching times of polarized light of  order ms.

\section{Conclusions}
We measured the time dependence of nuclear polarization  in a Ne gas cell. This was achieved by measuring the $\beta$-decay asymmetry of radioactive $^{20}$Na. The particles enter the gas cell as a fast beam where they stop and neutralize. The nuclear polarization is obtained using circular polarized laser light. The Na atoms should remain unbound to maintain nuclear polarization. Appreciable polarization can only be observed after dispensing natural Na in the gas cell. Keeping the cell at room temperature, where the vapor pressure for Na is negligible, means  that natural Na binds the contaminants in the cell that otherwise would bind $^{20}$Na. The long lifetime of the polarization shows that very few active contaminants remain. The maximum polarization is of order  50\%, indicating that only this fraction is available as atoms after the incoming beam has thermalized. 

We can describe the working of our gas cell using only three free parameters. However our description  remains qualitative. The question to what extent fast ions thermalize as atoms or ions can not be answered by this experiment. Improving the time resolution in switching between polarization directions will be necessary. In this way, polarized radioactive nuclei may provide an additional tool for studying plasma dynamics.

\section*{Acknowledgements} 
We thank the AGOR cyclotron staff for providing the beam and L. Huisman for technical support.
This research was financially supported by the “Stichting voor Fundamenteel
Onderzoek der Materie (FOM)” under Programme 114 (TRI$\mu$P) and ``FOM projectruimte''
08PR2636-1.

\end{document}